\documentclass[prd,aps,twocolumn,nofootinbib]{revtex4}
\usepackage{color}
\usepackage{graphicx}

\begin{document}
\title{Viscous damping  of chiral dynamos in the early universe}

\author{Andrii Neronov$^{1,2}$ and Dmitri Semikoz$^1$ }
\affiliation{$^1$Universit\'e de Paris,
CNRS, Astroparticule et Cosmologie, F-75013 Paris, France\\
$^2$Astronomy Department, University of Geneva, Ch. d'Ecogia 16, 1290, Versoix, Switzerland}

\begin{abstract}
Chiral dynamo converting asymmetry between right and left-handed leptons in the early universe into helical magnetic field  has been proposed as a possible cosmological magnetogenesis scenario. We show that this mechanism is strongly affected by viscous damping of primordial plasma motions excited by the dynamo. This effect modifies the expected range of strength and correlation length of the chiral dynamo field which could have survived till present epoch in the voids of the Large Scale Structure. We show the range of parameters of chiral dynamo field that may have survived in the voids  is still consistent with existing lower bounds on intergalactic magnetic field from gamma-ray observations, but only if the  right-left lepton asymmetry at the temperature $T\sim 80$~TeV is very high, close to the maximal possible value. 
\end{abstract}
\maketitle

%%%%%%%%%%%%%%%%%%%%%%%%%%%%%%%%%%%%%%%%%%%%%%%%%%
\section{Introduction}
%%%%%%%%%%%%%%%%%%%%%%%%%%%%%%%%%%%%%%%%%%%%%%%%%%

Magnetic fields found in all astronomical objects are  produced via dynamo action on weaker pre-existing fields \cite{kronberg,subramanian}. The weakest "seed" magnetic field which has given rise to the first dynamos could have been produced in the early universe, before the epoch of recombination.  Most of the cosmological magnetogenesis models  aiming at explanation of the origin of this seed field consider generation of magnetic fields at cosmological phase transitions, including the Electroweak and quark confinement phase transition, or Inflation  \cite{durrer13,subramanian}. 

A remarkable exception is the model proposed by Joyce and Shaposhnikov  \cite{joyce97} in which cosmological magnetic field is generated at the temperature $T\gtrsim 80$~TeV (i.e. much above the Electroweak phase transition temperature $T_{EW}\sim 100$~GeV). The hyper-magnetic field is generated via dynamo amplification of thermal fluctuations of the  $U(1)$ hyper-charge field. The dynamo is powered by the chiral asymmetry, which is a difference between the densities of right and left handed leptons. Such difference is destroyed by chirality flipping reactions as soon as the temperature drops below $T_R\simeq 80$~TeV,  \cite{campbell92,joyce97,80tev}, but in the temperature range $T>T_R$ the rate of chirality flipping reactions is smaller than the expansion rate of the universe and the right-left lepton asymmetry is approximately conserved for small enough values of the asymmetry parameter
\begin{equation}
\delta_R=\frac{n_R-n_L}{s}
\end{equation}
where $n_R,n_L$ are the densities of the right and left-handed leptons and $s=(2\pi^2/45)N_{eff}T^3$ is the entropy density, with $N_{eff}$ being the effective number of relativistic degrees of freedom. 

Ref. \cite{joyce97} has noticed that as soon s $\delta_R$ is above certain threshold value, anomalous coupling between chirality and hyper-magnetic helicity leads to an exponential growth of thermal fluctuations of hyper-magnetic field, up to the value in which magnetic helicity is in equipartition with chirality. The coupling is expressed in an additional term in the magnetic induction equation
\begin{equation}
\label{eq:induction0}
\frac{\partial \vec B}{\partial t}+\eta\vec\nabla\times  \left(\vec\nabla \times \vec B+4\mu \vec B\right)=0
\end{equation}
where $\eta$ is magnetic diffusivity, $\mu$ is the chemical potential  \cite{joyce97}
\begin{equation}
\mu=\frac{\alpha'}{\pi}\frac{2\pi^2}{45}\frac{783}{88}N_{eff}\delta_R T=\frac{\alpha'}{\pi} \frac{783}{88}\frac{(n_R-n_L)}{T^2}
\end{equation}
(relation valid for $\mu\ll T$) and  $\alpha'$ is the hyper-charge coupling.

\begin{figure}
\includegraphics[width=\linewidth]{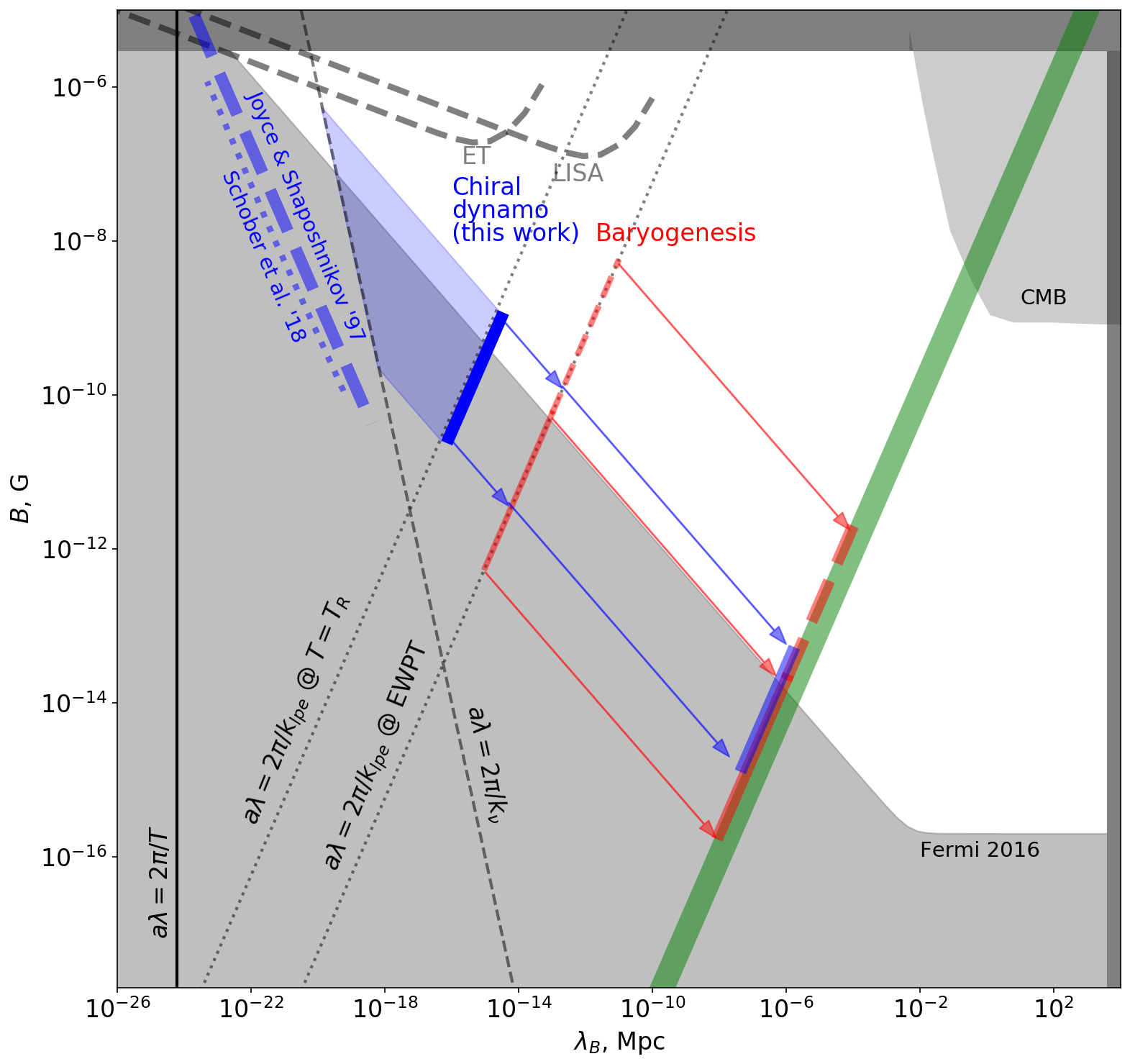}
\caption{Comparison of the estimates of the comoving strength and correlation length of the chiral dynamo magnetic field (this work, thick blue solid line and blue shaded range) with previous estimates from Refs. \cite{joyce97,schober18} and with the estimates of  the field necessary for successful baryogenesis (\cite{fujita,kamada16_1}, thick red solid and dashed lines). 
Light grey shading shows known limits on cosmological magnetic fields from gamma-ray observations  \cite{fermi} and CMB data \cite{planck}. Dashed grey lines shows the sensitivities of LISA and ET for MHD turbulence incuded gravitational wave backgorund \cite{neronov20}.  Darker grey shading shows theoretical limits.  Green band show the locus of the end points of cosmological evolution through the MHD turbulence inverse cascade \cite{banerjee,durrer13}.  Arrows show selected evolutionary paths.   Black dashed and dotted  lines show the thermal, viscous damping and largest processed eddy scales. }
\label{fig:exclusion}
\end{figure}

According to Ref. \cite{joyce97}, the chiral dynamo efficiently amplifies  magnetic field fluctuations with characteristic wavelengths $\lambda_\mu=2\pi/k_\mu=\pi/\mu$ and produces helical magnetic field with the strength $B_\mu\sim 3\times 10^2\mu T$. The estimate of parameters of the chiral dynamo magnetic field from Ref. \cite{joyce97} is shown in Fig. \ref{fig:exclusion} by the thick dashed blue line.

%%%%%%%%%%%%%%%%%%%%%%%%%%%%%%%%%%%%%%%%%%%%%%%%%%
%%%%%%%%%%%%%%%%%%%%%%%%%%%%%%%%%%%%%%%%%%%%%%%%%%

The helical magnetic field produced by the chiral dynamo subsequently evolves via development of turbulent cascade which preserves helicity \cite{banerjee,durrer13} and 
can ultimately produce relic magnetic fields surviving till present epoch. The predicted present day range of field strengths is interesting from the observational point of view  \cite{durrer13}. Starting from the moment of production, the field evolves respecting the conservation of helicity which, in terms of the comoving strength / length scale parameters, $\tilde B,\tilde \lambda$, is expressed as 
$\tilde B^2(t)\tilde \lambda(t)=const$ until the recombination epoch. The end points of the field evolution line up along the line $\tilde B\simeq 10[\tilde \lambda/1\mbox{ Mpc}]$~nG which corresponds to the largest processed eddy scale \cite{banerjee}.

The helical magnetic field that survives till the present epoch can be detected as intergalactic magnetic field (IGMF) using the $\gamma$-ray detection techniques \cite{kronberg,durrer13}. The lower bounds on IGMF are derived from non-observation of extended and delayed cascade $\gamma$-ray emission around extragalactic sources \cite{plaga95,neronov07,neronov10}. The analysis of 10-year data set of Fermi/LAT telescope \cite{fermi} has yielded a robust constraint on IGMF from  time delay of the cascade emission,  at the level $\sim 10^{-16}$~G for long correlation length fields and $\sim 10^{-14}$~G for the fields with correlation length about 1 pc (see Fig. \ref{fig:exclusion}). 

Strong enough chiral dynamo field might also be detectable through its stochastic gravitational wave backgorund excited through the magneto-hydro-dynamic (MHD) turbulence \cite{caprini09,pol20,neronov20}. Next generation gravitational wave detectors LISA \cite{lisa}, Einstein Telescope (ET) \cite{ET} have sensitivities for MHD turbulence driven gravitational wave background generated at the temperatures $T>100$~GeV. Fig. \ref{fig:exclusion} shows an estimate of sensitivities of LISA  and ET for the gravitational wave backgorund generated by magnetic fields with different strengths and correlation lengths estimated using the approach of Ref. \cite{neronov20}.  The strongest chiral dynamo field appears to be marginally detectable with LISA and ET.

In what follows we  show that previous treatments of the chiral dynamo, such as those of Refs. \cite{joyce97,schober18}  are incomplete in the sense that they ignore possible effects of excitation and damping of plasma motion by the magnetic field. Such plasma motions   damp magnetic field energy into heat. This leads to significant modification of the chiral dynamo action and changes the predictions for the field which might survive till present day in the voids of the Large Scale Structure. 

%%%%%%%%%%%%%%%%%%%%%%%%%%%%%%%%%%%%%%%%%%%%%%%%%%
\section{Chiral MHD equations}
%%%%%%%%%%%%%%%%%%%%%%%%%%%%%%%%%%%%%%%%%%%%%%%%%%

Plasma motion effects can be accounted in the framework of chiral Magneto-Hydro-Dynamics (MHD) in which the induction equation is modified to take into account possible non-zero plasma velocity \cite{boyarsky15,rogachevskii17}. The (hyper)-MHD equations can be conveniently expressed in the comoving coordinates through the derivatives with respect to the conformal time $\tilde t$ introduced as $d\tilde t=\int dt / a$, where $t$ is physical time and $a$ is the scale factor of the universe. Evolution of cosmological magnetic field is guided by the MHD equations which in the temperature  range $T\gg m_e$ are modified by the addition of the evolutionary equations for the chiral chemical potential \cite{boyarsky15,rogachevskii17,schober18}:
\begin{eqnarray}
\label{eq:rho}
&&\frac{\partial \tilde \rho}{\partial \tilde t}+\vec \nabla((\tilde \rho+\tilde p)\vec v)=0\\
\label{eq:v}
&&\frac{\partial \vec v}{\partial\tilde  t}+(\vec  v\vec\nabla)\vec v+
\frac{\vec v\partial \tilde p/\partial \tilde t}{(\tilde \rho+\tilde p)}+\frac{\vec\nabla \tilde p}{(\tilde \rho+\tilde p)} +\frac{ \tilde B\times (\vec\nabla \times  \tilde B)}{(\tilde \rho+\tilde p)}=\nonumber\\
&&\tilde\nu\left(\nabla^2 \vec v+\frac{1}{3}\vec\nabla(\vec\nabla \vec v)\right)\\
\label{eq:induction}
&&\frac{\partial \tilde B}{\partial \tilde t}-\vec\nabla\times( \vec v \times \tilde B)=\tilde \eta  \vec \nabla\times \left( \vec \nabla\times \tilde B+
4\tilde \mu\tilde B\right)\\
\label{eq:mu}
&&\frac{\partial \tilde \mu}{\partial \tilde t}+(\vec v\vec\nabla)\tilde \mu+\tilde \Gamma_f\tilde \mu=-\tilde \lambda\tilde \eta\left(\tilde B\cdot(\vec\nabla\times\tilde B)+4\tilde \mu \tilde B^2 \right)
\end{eqnarray}
Here $\tilde \nu$ is the shear viscosity, $\tilde \lambda$ is the  chiral feedback parameter and $\tilde \Gamma_f$ is the  rate of chirality flipping reactions. The comoving quantities denoted by tilde are related to their physical counterparts as follows: 
\begin{eqnarray}
&&\tilde B=a^2B,\  \tilde \rho=a^4\rho,\  \tilde p=a^4 p,\  \tilde \nu=\frac{\nu}{a}, \nonumber\\
&&\tilde\eta=\frac{\eta}{a},\ \tilde\mu =a \mu, \ \tilde\lambda=\frac{\lambda}{a^2},\ \tilde \Gamma_f=a\Gamma_f.
\end{eqnarray}

The magnetic diffusivity, the chirality flip rate and the chiral feedback parameter have are functions of  $T$. In the temperature range above the electroweak phase transition $T\gg 100$~GeV $\lambda$ scales as \cite{joyce97}
\begin{equation}
\lambda\simeq
\frac{783\alpha'^2}{88 \pi^2 T^2}
\end{equation}
The chirality flip rate is 
\begin{equation}
\Gamma_f=
\frac{T_R}{M_0}T
\end{equation}
where $T_R\simeq 80$~TeV is the temperature at which the chirality flip rate is equal to the expansion rate of the universe, $\Gamma_f=H$ and $M_0=\sqrt{90/(8\pi^3 N_{eff} G_N)}\simeq 7\times 10^{17}$~GeV, with $G_N$ being the Newton's constant and $N_{eff}=106.75$ is the number of relativistic degrees of freedom  in the temperature range of interest, $T\gtrsim 100$~GeV. In this temperature range
the magnetic diffusivity is \cite{joyce97}
\begin{equation}
\eta=\frac{1}{\sigma}\simeq \frac{1}{10^2T}.
\end{equation}
 
 %%%%%%%%%%%%%%%%%%%%%%%%%%%%%%%
\section{Chiral dynamo}
%%%%%%%%%%%%%%%%%%%%%%%%%%%%%%%

In this section we restore the line of reasoning of Ref. \cite{joyce97}, starting from the full system of hyper-MHD equations, (\ref{eq:rho})-(\ref{eq:mu}). We consider initial conditions with non-zero constant $\mu$ and we are interested in the effect of $\mu$ onto hyper-magnetic field $B$. 

Non-zero $\mu$ provides an exponential growth term in the  induction equation (\ref{eq:induction}). the time scale of the  exponential growth of magnetic field modes with wavenumbers $\tilde k$ can be established based on the structure of the last term of the induction equation:   
\begin{equation}
\tilde \tau=\frac{1}{\tilde \eta\tilde k (4\tilde\mu - \tilde k)} 
\end{equation}
The maximal growth rate is achieved at the wavenumber \cite{joyce97}
\begin{equation}
\label{eq:knaive}
\tilde k_\mu=2\tilde \mu
\end{equation}
For this mode, the characteristic growth scale is
\begin{equation}
t_\mu=a\tilde\tau_\mu=\frac{a}{4\tilde\eta\tilde\mu^2}\simeq \frac{10^2T}{4\mu^2}
\end{equation}

Requirement that the growth time scale should be shorter than the Hubble time $t_H=M_0/T^2$ constrains $\mu/T$ to be \cite{joyce97}
\begin{equation}
\label{eq:deltamin}
\delta =\frac{\mu}{T}\gtrsim 5\left(\frac{T}{M_0}\right)^{1/2}\simeq 2\times 10^{-6}\left[\frac{T}{T_R}\right]^{1/2}
\end{equation}
If the left-right asymmetry is generated at the temperature lower than $T_R$, the time available for the dynamo action is $\Gamma_f^{-1}$, rather than the Hubble time. In this case the requirement on successful dynamo action becomes $t_\mu<\Gamma_f^{-1}$ which gives
\begin{equation}
    \delta>\sqrt{\frac{T_R}{M_0}}\simeq 2\times 10^{-6}
\end{equation}
with no dependence on temperature.

Magnetic field strength can eventually become high enough so that it starts to produce back reaction on $\mu$. This back reaction occurs through the terms on the right hand side of Eq. (\ref{eq:mu}). An estimate of the field strength at which the backreaction  becomes important can be obtained based on the conservation of total chirality (particle chirality plus magnetic helicity)on the time scale $H^{-1}$ if $T>T_R$ and $\Gamma_f^{-1}$ if $T<T_R$ \cite{joyce97}
\begin{equation}
\frac{B^2}{k_\mu}\sim \frac{88}{783}\frac{2\pi^2 \mu T^2}{\alpha'^2}
\end{equation}
This gives 
\begin{equation}
\label{eq:Bnaive}
B_\mu=\sqrt{\frac{88}{783} }\frac{2\pi}{\alpha'}T^2\delta\simeq 3\times 10^2T^2\delta, 
\end{equation}
consistently with the original conclusion of Ref. \cite{joyce97}. Converting the estimates (\ref{eq:Bnaive}) and (\ref{eq:knaive}) to the comoving coordinates we find the estimates of $\tilde B$, $\tilde \lambda_B=2\pi/\tilde k_\mu$ shown in Fig. \ref{fig:exclusion} by the dashed thick blue line.

%%%%%%%%%%%%%%%%%%%%%%%%%%%%%%%%%%%%%%%%%%%
\section{Magnetic field and plasma motions}
%%%%%%%%%%%%%%%%%%%%%%%%%%%%%%%%%%%%%%%%%%%

The estimate (\ref{eq:Bnaive}) does not take into account a possibility that magnetic field amplification process is affected by plasma motion effects, such as turbulent cascade or viscous damping. To see if these effects are important,  we insepct Eq. (\ref{eq:v}). 
Magnetic field provides a source term for the velocity, which tends ultimately to reach the Alfven velocity 
\begin{equation}
v_A\equiv\sqrt{\frac{\tilde B^2}{(\tilde  \rho+\tilde p)}}\sim \sqrt{\frac{30 }{\pi^2 N_{eff} }}\frac{B}{T^2}\simeq 0.15\frac{B}{T^2}
\end{equation}
if $B=B_\mu$, then 
\begin{equation}
v_A\simeq 50\delta
\end{equation}
Making an order-of-magnitude comparison of relevant terns in the Navier-Stokes equation (\ref{eq:v}) one can find that in the absence of viscous damping this velocity scale can be reached on the time scale $\tau_v$ found from relation
$v_A/\tilde \tau_v\sim \tilde k_\mu v_A^2$
which gives $t_v=a\tilde\tau_v= a/(v_A\tilde k_\mu)\simeq T/(10^2\mu^2)$.
This time scale is much shorter than the Hubble time:
\begin{equation}
\frac{t_v}{t_H}\simeq \frac{10^{-15}}{\delta^2}\left[\frac{T}{T_R}\right]
\end{equation}
even for the smallest possible $\delta\sim 10^{-6}$. Thus, amplification of $B$ by the chiral dynamo inevitably excites plasma motions. 

Plasma motion provides back-reaction on the magnetic field through the mode coupling term, which is the second term in the induction equation (\ref{eq:induction}). Turbulence re-distributes the power injected at the scale $k_\mu$ across all scales up to the "largest processed eddy" scale with wavenumber  \cite{brandenburg,brandenburg1}
\begin{equation}
\label{eq:lpe}
k_{lpe}\equiv\frac{H}{v_A}\simeq \frac{T^2}{v_A M_0}\simeq \frac{6T^4}{BM_0}
\end{equation}
if $T\ge T_R$ and to
\begin{equation}
\label{eq:lpe1}
k_{lpe}\equiv\frac{\Gamma_f}{v_A}\simeq \frac{T_R T}{v_A M_0}\simeq \frac{6T_RT^3}{BM_0}
\end{equation}
in the temperature range $T<T_R$.

The modes with $k\sim k_\mu$ are affected by the turbulence as soon as $k_{lpe}<k_\mu$. This condition is satisfied if 
\begin{equation}
\label{eq:limit}
B> 3\times 10^{-13}T^2\delta^{-1}
\left\{\begin{array}{ll}
\left[T/T_R\right],& T>T_R\\
1,& T<T_R
\end{array}
\right.
\end{equation}

In the absence of viscous damping (see below), the inverse cascade transfers the magnetic field power toward the largest processed eddy scale while preserving the magnetic helicity, so that the relation
\begin{equation}
\frac{B_{lpe}^2}{k_{lpe}}\sim \frac{2\pi^2}{\alpha'^2}\frac{88}{783}\mu T^2
\end{equation}
holds. This provides an estimate of the magnetic field
\begin{equation}
\label{eq:Blpe}
B_{lpe}\simeq 3\times 10^{-3}T^2\delta^{1/3}
\left\{\begin{array}{ll}
\left[T/T_R\right]^{1/3},& T>T_R\\
1, & T<T_R
\end{array}
\right.
\end{equation}
which is much weaker than that in the pure chiral dynamo setup (\ref{eq:Bnaive}). The characteristic wavenumber range of this weaker field is
\begin{equation}
\label{eq:lpe2}
k_{lpe}\simeq 2\times  10^{-10}T\delta^{-1/3}
\left\{\begin{array}{ll}
\left[T/T_R\right]^{2/3}, & T>T_R\\
1, & T<T_R.
\end{array}
\right.
\end{equation}

Comparing the limit (\ref{eq:limit}) to the estimate of $B_{lpe}$ we find that the constraint is satisfied as soon as
\begin{equation}
\delta> 2\times 10^{-8}
\left\{\begin{array}{ll}
\left[T/T_R\right]^{1/2}, & T>T_R\\
1, & T<T_R,
\end{array}
\right.
\end{equation}
which is always the case if the chiral dynamo is efficient. 

Thus, the effects of plasma motion cannot be ignored. Ref. \cite{schober18} has studied modifications of the chiral dynamo magnetic field estimates due to the plasma turbulence  for the temperature range down to $T\sim 100$~GeV. Estimates of the field parameters found in Ref. \cite{schober18} is shown by the blue thick dotted line in Fig. \ref{fig:exclusion}. Numerical simulaitons of Ref. \cite{schober18} were performed under assumption $\Gamma_f=0$ on time scales much shorter than the Hubble time. In this setup the turbulence does not have time to develop on long wavelength scales $\lambda_{lpe}=2\pi/k_{lpe}$. The simulations also assumed small Prandtl numbers $Pr\sim 1$, potentially under-estimating the influence of viscosity on plasma motions. 

%%%%%%%%%%%%%%%%%%%%%%%%%%%%%%%%%%%%%%%%%%%
\section{Viscous damping of  plasma motions}
%%%%%%%%%%%%%%%%%%%%%%%%%%%%%%%%%%%%%%%%%%%

The Alfven velocity scale can be reached only if the viscous damping term in the Navier-Stokes equation (\ref{eq:v}) is negligible at the relevant distance / wavenumber scales, including the forcing scale $k_\mu$. This condition can be verified through the order-of-magnitude comparison of the source and viscous damping terms. Order-of-magnitude comparison of  the second term on the left of the equation (\ref{eq:v}) and the right-hand side indicates that the viscous damping is important for the modes with wavenumber $\tilde k$ if $\tilde \nu \tilde k^2 v\gtrsim \tilde k v^2$, i.e. if 
\begin{equation}
\label{eq:estimate}
k\gtrsim \frac{v_k}{\nu} 
\end{equation}
where $v_k$ is the characteristic velocity scale of the modes with wavenumber $\tilde k$. This velocity scale can be estimated knowing that turbulence forms Kolmogorov power spectrum in the wavenumber range $\tilde k>\tilde k_{lpe}$

One can  introduce  $v_k$ as \cite{durrer13}
\begin{equation}
v_k=\sqrt{\frac{P_K(\tilde k)\tilde k^3}{2\pi^2}}
%B_k=\sqrt{\frac{P_B(k)k^3}{2\pi^2}}; 
\end{equation}
where $P_K(k)$ is the Fourier power spectrum of velocity field. The symptotic behaviour of $P_k$ at small $k$ is 
$P_K(k)\propto k^0$. Large $k$ asymptotic is   $P_K(k)\propto k^{-\alpha_K}$ with $\alpha_K=11/3$ at large $k$. Thus, the velocity $v_k$ scales as 
\begin{equation}
v_k(\tilde k)\propto v_A\left(\frac{\tilde k_{lpe}}{\tilde k}\right)^{1/3}
\end{equation} 
Substituting this expression into (\ref{eq:estimate}) we find that viscous damping is important for the modes with $k>k_\nu$, where
$$
k_\nu^{4/3}=\frac{v_A k_{lpe}^{1/3}}{\nu}\simeq \frac{v_A k_{lpe}^{1/3}T}{20}
$$
where we have used the estimate of viscosity \cite{arnold00,durrer13}
\begin{equation}
\nu\sim \frac{\lambda_{mfp}}{5}\simeq \frac{20}{T}
\end{equation}
where $\lambda_{mfp}$ is the mean free path of the least coupled particles, which are the  right-handed electrons  in the case of plasma at temperature $T\gg 100$~GeV. The requirement $k_\nu>k_\mu$ (no viscous damping on the magnetic field forcing scale) translates to
\begin{equation}
\delta<3\times 10^{-8}
\left\{\begin{array}{ll}
\left[T/T_R\right]^{1/2}, &T>T_R\\
1, & T<T_R.
\end{array}\right.
\end{equation}
Viscous damping is always important on the scale $k_\mu$ for $\delta\gtrsim 2\times 10^{-6}$, in the range considered in the context of the chiral dynamo.  

Damping of the dynamo action by viscosity is less important for the modes with smaller wavenumbers, $k\ll \mu$. Even though the growth rate of these modes is much lower than that of the modes with $k\sim \mu$, chiral dynamo acting on these long-wavelength modes can still be efficient on the time scale comparable to the Hubble time if $T>T_R$ and to the chirality flip time if $T<T_R$. Comparing the characteristic time scale of viscous damping 
$\tilde t_\nu(\tilde k)=(\tilde \nu \tilde k^2)^{-1}$
with the growth time scale due to the chiral dynamo
$\tilde t_\mu(\tilde k)=(4\tilde \eta \tilde k\tilde\mu)^{-1}$
 one can find that the necessary condition for the mode growth is 
 \begin{equation}
 k<k_{\mu}^*=\frac{4\tilde \eta \tilde\mu}{\tilde \nu}\simeq 3\times 10^{-3}\mu
 \end{equation}
 The growth time scale of these modes is $ t_\mu=10^4T/\mu^2$.
Comparing this time scale to the Hubble time or the chirality flip time one finds that the condition for efficient dynamo with account of the viscous damping of the short wavelength modes is 
\begin{equation}
\delta> 3\times 10^{-5}\left\{\begin{array}{ll}
\left[T/T_R\right]^{1/2},& T>T_R\\
1, &T<T_R.
\end{array}\right.
\end{equation}
Thus, account of viscosity effects modifies  the naive estimate of the limit of the chiral dynamo efficiency (\ref{eq:deltamin}) by more than an order of magnitude. It also modifies the estimate of the wavenumber range in which the dynamo operates. 

Shift of the turbulence forcing scale from $k\sim \mu$ to $k\sim 3\times 10^{-3}\mu$ does not affect processing of the largest processed eddy modes by turbulence. In any case, the inverse cascade transfers the power toward  $k\sim k_{lpe}$, given by Eq. (\ref{eq:lpe}), because in any case, $k_{lpe}\ll k_\mu^*$. The main effect of the turbulence and viscous damping is in the shrinking of the range of $\delta$ in which the chiral dynamo can be efficient.  

The estimates of Eqs.  (\ref{eq:Blpe}), (\ref{eq:lpe2}) can be converted in the comoving frame 
\begin{eqnarray}
\label{eq:lpe1}
&&\tilde B_{lpe}\simeq 1\delta^{1/3}\left[\frac{N_{eff}}{10^2}\right]^{-2/3}\left[\frac{T}{T_R}\right]^{1/3}\mbox{ nG}\nonumber\\
&& \tilde \lambda_B=10^{10}\delta^{1/3}\left[\frac{N_{eff}}{10^2}\right]^{1/3}\left[\frac{T_R}{T}\right]^{2/3}\mbox{ cm}
\end{eqnarray}
for $T>T_R$. If $T<T_R$, this estimate is still valid upon substitution $T\rightarrow T_R$.
The range of possible values of $\tilde B_{lpe},\tilde \lambda_B$ for $\delta>3\times 10^{-5}$ is shown by the thick solid blue line in Fig. \ref{fig:exclusion}. Blue shaded region shows a range of possible peak values of $\tilde B,\tilde \lambda_B$ that can potentially be reached on time scales shorter than $H^{-1}$ ($T>T_R$) or $\Gamma_f^{-1}$ ($T<T_R$). 

The viscous damping effect has not been taken into account in the analysis of Refs. \cite{joyce97,schober18}. Fig. \ref{fig:exclusion} shows that in both cases the field parameter estimates fall into the sector of $\tilde B, \tilde \lambda_B$ parameter space where viscous damping cannot be ignored (to the left of the black dashed line). Our estimates show that account of the viscous damping changes the expected initial parameters of the field.

%%%%%%%%%%%%%%%%%%%%%%%
\section{Discussion and conclusions}
%%%%%%%%%%%%%%%%%%%%%%%

We have found that the effect of viscous damping of plasma motions strongly affects the estimates of parameters of cosmological magnetic fields produced by chiral dynamos operating in the Early Universe. This effect has not been taken into account in previous calculations of Ref. \cite{joyce97,schober18}. The viscous damping shifts the characteristic wavenumbers of modes experiencing the chiral dynamo effect toward lower values thus reducing the growth rate of the dynamo. This reduces the range of possible initial chiral asymmetry $\delta$ in which the dynamo can produce sizeable primordial magnetic field. Besides, viscous damping reduces the peak values of the field, moving them out of the range which can be probed through the stochastic gravitational wave backgorund measurements with LISA and ET (see Fig. \ref{fig:exclusion}).  

Comparing the expected range of magnetic field strength with the lower bound on IGMF from gamma-ray observations, we find that the IGMF present in the voids of the LSS can well be the field originating from the chiral dynamo mechanism if 
\begin{equation}
\label{eq:003}
\delta\gtrsim 0.03.
\end{equation}
This condition does not rule out the possibility of existence of primordial magnetic field generated by dynamos driven by lower $3\times 10^{-5}<\delta<0.03$, it just indicates that weaker IGMF produced by such dynamos is not detectable with $\gamma$-ray telescopes (and hence, such chiral dynamo models cannot be tested with any known observational technique). The range (\ref{eq:003}) is at the limit of applicability of assumption $\mu\ll T$ implicitly used in the treatment presented above. This suggests that detailed comparison of chiral dynamo model predictions with observational data needs to be re-assessed in $\mu\lesssim T$ regime.

The hyper-magnetic field produced by the chiral dynamo can play an important role in cosmology. Turbulent decay of this field leads to the helical magnetic field at the epoch of Electroweak phase transition which is strong enough to provide baryogenesis via conversion of helicity into baryon asymmetry \cite{giovaninni,semikoz_v,fujita,kamada16,kamada16_1}. The modified estimates of the chiral dynamo field strength and correlation length with account of turbulence and viscosity are within the range of the estimates of the field parameters necessary for production of the observed level of baryon-to-photon ratio in the Universe, $\sim 10^{-10}$, which, in terms of the present-day comoving magnetic field parameters range between $10^{-14}$~G$<\tilde B<10^{-12}$~G \cite{fujita}
and $10^{-16}$~G$<\tilde B<10^{-14}$~G \cite{kamada16_1}, depending on assumptions about uncertain temperature dependence of of the hyper-magnetic-to-magnetic field transition which occurs in the temperature range $T\sim 100$~GeV. 

Magnetic fields with large enough correlation length evolve through turbulent decay. During the decay, the field correlation length is always equal to the largest processed eddy scale $l_{lpe}\sim 1/k_{lpe}\sim B M_0/T^4$ (see Eq. (\ref{eq:lpe}) \cite{banerjee,durrer13}.  The growth of the correlation length is accompanied by the decrease of the field strength so that the helicity is preserved. As a result, the field evolution follows the path $\tilde B^2\tilde \lambda_B=const$. Selected evolutionary paths are shown by red and blue arrows in Fig. \ref{fig:exclusion}. One can see from Fig. \ref{fig:exclusion} that evolution of the chiral dynamo magnetic field in the temperature range from $T_R$ down to the electroweak phase transition temperature leads to the helical magnetic field with parameters compatible with those of the field required for the  successful baryogenesis. 

An observational test of this attractive "baryogenesis $+$ magnetogenesis" scenario is possible through the detection of the relic helical magnetic field in the present day universe. Green band in Fig. \ref{fig:exclusion} shows the locus of the end points of cosmological evolution of magnetic field at the epoch of recombination \cite{banerjee}. Grey lower bound on the relic cosmological magnetic field shown in the Figure is derived from the gamma-ray data, specifically from non-observation of delayed cascade emission in the signal of extragalactic gamma-ray sources by Fermi/LAT telescope \cite{plaga95,neronov07,neronov09,neronov10,fermi}. 

Only the strongest chiral dynamo field for $\delta\gtrsim 0.03$ can be still revealed by the gamma-ray observations. This requires significant coordinated observational effort of long-term monitoring of distant gamma-ray sources in multi-TeV band, on decade time scales, combined with long-term monitoring in multi-GeV band. The TeV band monitoring of large number of extragalactic gamma-ray sources is challenging but possible with the planned Cherenkov Telescope Array (CTA), which will probably have enough sensitivity to take "snapshot" measurements of hard gamma-ray spectrum active galactic nuclei  (AGN) on regular basis. Monitoring of the sources in the multi-GeV band is possible with space-based telescopes and currently Fermi/LAT telescope \cite{atwood09} provides such monitoring. HERD telescope \cite{herd} planned for launch in 2024 will have acceptance similar to Fermi/LAT and will be able to extend the long-term monitoring of AGN started by Fermi/LAT on multi-decade time span. Combination of CTA, Fermi/LAT and HERD data can provide and improvement of sensitivity reach of the gamma-ray technique needed for the full exploration of the parameter range of  magnetic field  from the chiral dynamo and possibly assuring the cosmological baryogenesis. 

\section*{Acknowledgement}

This work is supported by French National Research Agency project MMUniverse (ANR-19-CE31-0020). The authors are grateful to M.Shaposhnikov, O.Ruchayskiy, J.Schober and V.Semikoz for fruitful discussions.

\bibliography{references.bib}

\end{document}